\documentclass[11pt,a4paper]{article}
\usepackage{jheppub}

\usepackage{amsmath, amsfonts, amsthm, amssymb, graphicx}
\usepackage[dvipsnames]{xcolor}
\usepackage{epstopdf}
 \usepackage{slashed}
\usepackage{changepage}

\usepackage{pifont}

\usepackage{caption}
\def\be{\begin{equation}}
\def\ee{\end{equation}}
\def\ba{\begin{eqnarray}}
\def\ea{\end{eqnarray}}

\def\bdm{\begin{displaymath}}
\def\edm{\end{displaymath}}

\def\bq{\begin{quote}}
\def\eq{\end{quote}}

\def\del{\partial}
\def\ltap{\ \raise.3ex\hbox{$<$\kern-.75em\lower1ex\hbox{$\sim$}}\ }
\def\gtap{\ \raise.3ex\hbox{$>$\kern-.75em\lower1ex\hbox{$\sim$}}\ }
\def\gl{\ \raise.5ex\hbox{$>$}\kern-.8em\lower.5ex\hbox{$<$}\ }
\def\roughly#1{\raise.3ex\hbox{$#1$\kern-.75em\lower1ex\hbox{$\sim$}}}

 at 10truept

\newcommand{\N}{\mathcal{N}}

\newcommand{\beq}{\begin{equation}}
\newcommand{\eeq}{\end{equation}}
\newcommand{\bea}{\begin{eqnarray}}
\newcommand{\eea}{\end{eqnarray}}
\newcommand{\beqa}{\begin{eqnarray}}
\newcommand{\eeqa}{\end{eqnarray}}

\def \del {\partial}

\def\be{\begin{equation}}
\def\ee{\end{equation}}

\usepackage[dvipsnames]{xcolor}

\title{The cosmological constant is probably still zero}

\author[1,2]{Yang Liu,} 
\author[1,2]{Antonio Padilla} 
\author[3,4]{and Francisco G. Pedro}

 \affiliation[1]{School of Physics and Astronomy, University of Nottingham, University Park, Nottingham NG7 2RD, United Kingdom}
\affiliation[2]{Nottingham Centre of Gravity, University of Nottingham, Nottingham NG7 2RD, UK}
 \affiliation[3]{Dipartimento di Fisica e Astronomia, Universit\`a di Bologna, via Irnerio 46, 40126 Bologna, Italy}
\affiliation[4]{INFN, Sezione di Bologna, viale Berti Pichat 6/2, 40127 Bologna, Italy}

\emailAdd{yang.liu@nottingham.ac.uk}
\emailAdd{antonio.padilla@nottingham.ac.uk}
\emailAdd{francisco.soares@unibo.it}

\abstract{We consider a wide class of four-dimensional effective field theories in which gravity is coupled to multiple four-forms and their dual scalar fields, with membrane sources charged under the corresponding three-form potentials. Four-form flux, quantised in units of the membrane charges, generically generates a landscape of vacua with a range of values for the cosmological constant that is scanned through membrane nucleation. We list various ways in which the landscape can be made sufficiently dense to be compatible with observations of the current vacuum without running into the empty universe problem. Further, we establish the general criteria required to ensure the absolute stability of the Minkowski vacuum under membrane nucleation and the longevity of those vacua that are parametrically close by. This selects the current vacuum on probabilistic grounds and can even be applied in the classic model of Bousso and Polchinski, albeit with some mild violation of the membrane weak gravity conjecture. We present other models where the membrane weak gravity conjecture is not violated but where the same probabilistic methods can be used to tackle the cosmological constant problem.   
} 

\begin{document}
\maketitle
\section{Introduction}
The cosmological constant problem has haunted theoretical physics ever since Pauli calculated the effect of the zero-point energy of the electron on the curvature of spacetime, declaring that the universe would ``not even reach to the moon" \cite{Enz}. The situation hasn't really improved in the century that followed. When we apply standard quantum field theory methods, radiative corrections to the vacuum energy are extremely sensitive to ultra-violet physics,  scaling like the fourth power of the cut-off.  Vacuum energy gravitates just like a cosmological constant presenting a major issue for cosmology. The observed value of the cosmological constant lies sixty orders of magnitude below the expected value predicted from vacuum energy calculations with a TeV cut-off,  set by the the scale of modern collider experiments. If we push the cut-off all the way up to the Planck scale, the discrepancy extends to 120 orders of magnitude.  For reviews of the cosmological constant problem, see \cite{Weinberg:1988cp, Polchinski:2006gy, Burgess:2013ara,Padilla:2015aaa}.

Recently, Kaloper \cite{Kaloper:2022oqv} and later, Kaloper and Westphal \cite{Kaloper:2022jpv} have constructed a simple dynamical model in which the cosmological constant is relaxed to near zero through the nucleation of membranes, charged under a pair of three-forms.  Membrane nucleation triggers a jump in the effective cosmological constant that is controlled by the corresponding charges. Crucially, these charges are not assumed to be exponentially small. Instead, they are assumed to have an irrational ratio, which guarantees a dense landscape of vacua, including those whose cosmological constant is close to zero. Given a more  modest upper bound on the charge (which may be in some tension with the weak gravity conjecture \cite{Arkani-Hamed:2006emk,Ibanez:2015fcv})  transitions between vacua are rapid, at least until the system finds a vacuum with a small positive cosmological constant, at which point, the vacuum is very long lived.  This latter result has its roots in  Hawking's famous solution to the cosmological constant problem in which he computes the Euclidean action for a landscape of vacua and argues that the most probable configuration  is the one with the smallest absolute curvature  \cite{Hawking:1984hk}.

Dynamical neutralization of the cosmological constant via membrane nucleation was first explored by Brown and Teitelboim \cite{Brown:1987dd,Brown:1988kg}, although there the membrane charge is assumed to be extremely small. As a result, the transition from a very high energy de Sitter vacuum to near Minkowski  occurs slowly via a series of small steps, resulting in the so-called empty universe problem \cite{Abbott:1984qf}.  Thanks to the irrational charge ratio in \cite{Kaloper:2022oqv,Kaloper:2022jpv}  the transition from a very high energy de Sitter vacuum to near Minkowski can occur quickly via a single membrane nucleation and the empty universe problem is said to be avoided. 

The simple set-up presented in \cite{Kaloper:2022jpv} is a clever generalisation of Henneaux and Teitelboim's  covariant formulation of unimodular gravity \cite{Henneaux:1989zc}, with a pair of four-form field strengths that enter only through a bilinear mixing with a dual scalar field,  and charged membranes (for a review of unimodular gravity, see \cite{Padilla:2014yea}).  Although the model is phenomenologically interesting, its embedding within fundamental theory is likely to prove challenging. However, in this paper we will show that key aspects of the relaxation mechanism apply far more generally, increasing the likelihood that it could apply in some of the four-dimensional effective theories  obtained from string theory compactifications. Remarkably, this is true of the famous Bousso-Polchinski set-up \cite{Bousso:2000xa}, offering a probabilistic explanation for the small observed value of the cosmological constant instead of an anthropic one.  

 We consider a very general set-up  involving Einstein-Hilbert gravity coupled to three-forms fields and their dual scalars, with membrane sources charged under the three-forms.   This includes the Kaloper-Westphal model \cite{Kaloper:2022jpv} as a special case but also other effective theories including those expected to appear in compactifications of string theory, such as the Bousso-Polchinski set-up \cite{Bousso:2000xa}.  A key feature of the entirely family of models is that in vacuum, the scalars and the three-forms always gravitate like a cosmological constant, even in the presence of a non-trivial flux. The flux of the four-field strength is quantised, giving rise to a landscape of vacua. To have any hope of solving the cosmological constant problem, this landscape must be sufficiently dense in and around Minkowski space in order to admit vacua that match observations.

Tunnelling between vacua is achieved through membrane nucleation. This is described by instantons, solutions to the Euclidean equations of motion with vacua of different vacuum energies, separated by charged membranes. The geometry of the instantons depends on the curvature of the corresponding vacua and  the  membrane tensions. Many configurations are physically irrelevant because   the membranes would have to have  negative tension or because the corresponding tunnelling rate is infinitely suppressed.  As in \cite{Kaloper:2022oqv,Kaloper:2022jpv}, those instanton solutions that are physically relevant can be dissected even further. In our general setting, this dissection depends on the value of a parameter $X=4 M_{pl}^4 \Delta k^2/T^2$, where $\Delta k^2$ is the jump in vacuum curvature and $T$ is the  membrane tension. We consider several example theories where the bounds on this parameter can be related to bounds on other microscopic parameters  such as the membrane charges. This is important since the value of $X$ not only controls which instanton geometries are allowed but also the corresponding tunnelling rates. In particular, when $|X|<1$, any near Minkowski vacua become very long lived and robust against the nucleation of anti-de Sitter bubbles.  Thus, if you find yourself in a near Minkowski vacuum, chances are you will stay there. 

Of course, by itself, this is not enough to solve the cosmological constant problem.  One must show that we avoid the so-called empty universe problem. This occurs when the universe descends to the current vacuum incrementally, via a slow cascade of de Sitter vacua, the curvature  jumping by a small amount each time, forever diluting away any matter excitations.
 The density of the landscape and the abundance of dangerous de Sitter vacua render this a real possibility in generic theories of this type. However, through a detailed study of the kinematics and an analysis of tunnelling rates we show that the empty universe problem can often be avoided: one is able to transition quickly from a high scale de Sitter vacuum to near Minkowski via a single transition.  All of this suggests a wide-ranging mechanism for solving the cosmological constant problem, inspired by, but extending far beyond, the original proposal of \cite{Kaloper:2022oqv,Kaloper:2022jpv}. 

The rest of this paper is organised as follows: in section \ref{sec:setup}, we present the generalised set-up along with the corresponding field equations and junction conditions. Since they will play an important role in the evaluation of tunnelling rates later in the paper, we dwell a little on the choice of boundary conditions and corresponding boundary terms. In section \ref{sec:vacua}, we present the landscape of Lorentzian vacuum solutions. In section \ref{sec:rates}, we Wick rotate to Euclidean signature and solve the Euclidean field equations to find the corresponding instanton solutions and compute the corresponding transition rates. In section \ref{analysis},  we demonstrate the role of the parameter $X$ in controlling the stability of near Minkowski vacua, protecting them from decay into anti de Sitter. We also discuss conditions to avoid the empty universe problem. In section \ref{sec:examples}, we run through four different models, including Brown-Teitelboim, Bousso-Polchinski, Kaloper-Westphal,  and a fourth model including a single three-form (with standard kinetic term) and its dual scalar.  In the latter three cases, we show how the parameters of the theory can be chosen to successfully implement the mechanism for solving the cosmological constant problem originally proposed in \cite{Kaloper:2022oqv}. In section \ref{sec:conc}, we conclude. 

\section{The generalised set-up} \label{sec:setup}
We begin  with a general four-dimensional effective theory on a manifold, $\mathcal{M}$, with a dynamical metric $g_{\mu\nu}$ and a family of three-form fields, $A^i$,  and dual scalars $\phi_i$, 
\begin{multline} \label{Lorentzianaction}
S =\int_\mathcal{M} d^4 x\sqrt{|g|} \left[\frac{M_{pl}^2}{2}  R-\frac12 \omega^{ij}(\phi) \nabla_\mu \phi_i \nabla^\mu  \phi_j-V(\phi) \right] \\ +
\int_\mathcal{M} \left[ -\frac{1}{2}Z_{ij} (\phi)F^i\wedge \star F^j+\sigma_i(\phi) F^i \right]
 +S_\text{boundary}+S_\text{membranes} 
\end{multline}
      where $R$ is the Ricci  scalar. The four-form field strengths are given in terms of the three-form fields $F^i=dA^i$ and $\star$ denotes the Hodge star operator on the manifold $\mathcal{M}$. In components, the three-form is written as $A^i=\frac{1}{3!} A^i_{\mu\nu\alpha} dx^\mu\wedge dx^\nu \wedge dx^\alpha$ and the corresponding field strength as $F^i=\frac{1}{4!} F^i_{\mu\nu\alpha\beta} dx^\mu\wedge dx^\nu \wedge dx^\alpha\wedge dx^\beta$ where $F^i_{\mu\nu\alpha\beta}=4 \del_{[\mu}A^i_{\nu\alpha\beta]}$. When acting on the four-form, the Hodge star operator yields $\star F^i=\epsilon_{\mu\nu\alpha\beta} F^{i \mu\nu\alpha\beta}$, where $\epsilon_{\mu\nu\alpha\beta}$ is the Levi-Civita tensor  on the manifold.

The action is also equipped with boundary terms which depend on the choice of boundary conditions. These are integrals over the boundary, $\partial \mathcal{M}$, which we take to be a co-dimension one surface described by the embedding $x^\mu=X^\mu(\xi^a)$. The induced metric and the pullback of the three-forms on the boundary are given respectively by $\gamma_{ab}=g_{\mu\nu}X^\mu_{, a}X^\nu_{, b}$ and $\alpha^i=\frac{1}{3!} A^i_{\mu\nu \alpha} X^\mu_{, a}X^\nu_{, b}X^\alpha_{, c} d\xi^a \wedge d \xi^b \wedge d \xi^c$, where $X^\mu_{, a}=\del X^\mu/\del \xi^a$ are the boundary tangent vectors.  We also define the extrinsic curvature of the boundary, $K_{ab}=\frac12 \mathcal{L}_n \gamma_{ab}$, as the Lie derivative of the induced metric with respect to the outward pointing unit normal, $n^\mu$. If $K=\gamma^{ab} K_{ab}$ is the trace of the extrinsic curvature on the boundary, the boundary action is given by
\be
S_\text{boundary}=M_{pl}^2 \int_{\partial \mathcal{M}}d^3 \xi \sqrt{|\gamma|}   K  -\int_{\partial \mathcal{M}} \mu p^i \phi_i  +\lambda \chi_i\alpha^i 
\ee     
where we define ``conjugate momenta", 
\be
p^i=-d^3 \xi \sqrt{|\gamma|} \omega^{ij} n^\mu \nabla_\mu \phi_j,\quad \chi_i= \sigma_i-Z_{ij}( \star F^j). 
\ee
The extrinsic curvature piece ensures that the action can be extremised under metric variations with Dirichlet boundary conditions, $\delta \gamma_{ab}=0$ \cite{Gibbons:1976ue}. Meanwhile, the parameter $\lambda$ allows us to interpolate between Dirichlet ($\lambda=0$) and Neumann ($\lambda=1$) boundary conditions on the three-forms, while the parameter $\mu$ allows us to interpolate between Dirichlet ($\mu=0$) and Neumann ($\mu=1$)  boundary conditions on the scalars. To see this, note that variation of the action with respect to all fields yields a boundary contribution \cite{Gibbons:1976ue, Duncan:1989ug},
\be
-\frac{M_{pl}^2}{2} \int _{\partial \mathcal{M}} d^3 \xi \sqrt{|\gamma|} (K^{ab} -K \gamma^{ab}) \delta \gamma_{ab} 
+\int_{\partial \mathcal{M}} (1-\lambda) \chi_i\delta \alpha^i -\lambda \delta \chi_i \alpha^i 
+  (1-\mu) p^i\delta\phi_i-\mu \delta p^i \phi_i 
\ee
that is required to vanish under the appropriate choice of boundary conditions.  

Finally we consider the membrane contributions. We can include contributions from membranes and anti-membranes, $\Sigma_I$,  charged under any of the three-forms, such that
\be \label{membact}
S_\text{membranes}=-\sum_I \left\{\eta^i_I q_i\ \int_{\Sigma_I} \alpha_I^i + |\eta^i_I| \tau_i \int_{\Sigma_I} d^3 \xi \sqrt{|\gamma_I}| \right\}.
\ee
Membranes charged under $A^i$ carry a fundamental charge $\pm q_i$ depending on whether they are branes or antibranes and tension $\tau_i$. In the action \eqref{membact},  $\eta^i_I=0, \pm 1$ depending on whether the membrane $\Sigma_I$ carries positive ($\eta^i_I= 1$), negative ($\eta^i_I=- 1$) or vanishing charge ($\eta^i_I=0$) under $A^i$. The induced metric and the pullback of the three-forms on $\Sigma_I$  are given  in a similar way to the boundary, by $\gamma_I{}_{ab}=g_{\mu\nu}X_I{}^\mu_{, a}X_I{}^\nu_{, b}$ and $\alpha_I^i=\frac{1}{3!} A^i_{\mu\nu \alpha} X_I{}^\mu_{, a}X_I{}^\nu_{, b}X_I{}^\alpha_{, c} d\xi^a \wedge d \xi^b \wedge d \xi^c$, where $X_I{}^\mu_{, a}=\del X_I{}^\mu/\del \xi^a$ are the  tangent vectors on $\Sigma_I$. Throughout this paper we will restrict attention to timelike membranes so that their unit normal, $n^\mu_I$ is spacelike.

Away from the membranes,  the field equations resulting from variation of the action \eqref{Lorentzianaction} with respect to the metric, three-forms and scalars are given by
\begin{eqnarray}
M_{pl}^2 G^{\mu\nu} &=&T^{\mu\nu}_F+T^{\mu\nu}_\phi  \\
\nabla_\mu \chi_i &=&0\\
\nabla_\mu \left( \omega^{kj} \nabla^\mu \phi_j\right) &=&V^{, k}+\frac12 \omega^{ij}{}^{,k} \nabla_\mu \phi_i \nabla^\mu  \phi_j-\left[\frac{1}{2}Z_{ij}{}^{, k}  (\star F^i) (\star F^j)-\sigma_{i}{}^{,k} (\star F^i )\right] \qquad 
\end{eqnarray}
where  $V^{,k}=\del V/\del\phi_k$ etc. The energy momentum tensors for the three-forms, the scalar and the membranes are \begin{eqnarray}
&& T^{\mu\nu}_F = \frac{Z_{ij}}{3!}\left( F^i{}^{\mu\alpha\beta\gamma}F^j{}^{\nu}{}_{\alpha\beta\gamma}-\frac18 g^{\mu\nu} F^i_{\alpha\beta\gamma\delta}F^j{}^{\alpha\beta\gamma\delta}\right)\qquad \\
&& T^{\mu\nu}_\phi  =  \omega^{ij}\left(\nabla^\mu \phi_i \nabla^\nu  \phi_j-\frac12 g^{\mu\nu} \nabla_\alpha \phi_i \nabla^\alpha  \phi_j\right) -Vg^{\mu\nu}.\qquad 
\end{eqnarray}
At the membranes, we can derive junction conditions which determine how the conjugate momenta of the system jump across the membranes.  The momentum conjugate to $\gamma_{I ab}$ is proportional to the following combination of the extrinsic curvature $\pi_I^{ab} =M_{pl}^2 (K_I^{ab}-K^I \gamma_I^{ab})$. As we pass over  $\Sigma_I$ this  jumps  in accordance with the Israel junction conditions \cite{Israel}
\be
M_{pl}^2 \Delta \pi_I^{ab}= -|\eta^i_I| \tau_i \gamma_I^{ab}.
\ee
The momentum conjugate to the three-forms also jumps
\be \label{chijump}
\Delta \chi_i=-\eta^i_I q_i \qquad  (\text{no sum over $i$})
\ee
while the scalar momenta remains continuous $\Delta p^i=0$.  The metric and the three-form potentials are assumed to be continuous at $ \Sigma_I$ in order for the membrane action to be well defined.  Note that the junction condition \eqref{chijump} is consistent with the momentum $\chi_i$ quantised in units of the fundamental membrane charge, $\chi_i=-N_i q_i$ (no sum)  where $N_i \in \mathbb{Z}$.

\section{Vacua} \label{sec:vacua}
We take vacua to be real Lorentzian solutions with constant scalars, four-forms of constant flux $\star F^i=c^i$, and a maximally symmetric  metric with constant curvature $k^2$, corresponding to   de Sitter ($k^2>0$),  Minkowski ($k^2=0$) or anti de Sitter ($k^2<0$) spacetime. Away from any membranes, the field equations imply that
\begin{eqnarray}
&& 3M_{pl}^2 k^2 =V+\frac{1}{2} Z_{ij}c^i c^j  \label{con1} \\
&& V^{, k}=\frac{1}{2}Z_{ij}{}^{, k}  c^i c^j-\sigma_{i}{}^{,k} c^i \label{con2}
\end{eqnarray}
while the conjugate momentum, 
\be
\chi_i = \sigma_i- Z_{ij} c^j  \label{con3}
\ee
is  locally constant.  At membranes, we get a jump in $\chi_i$  according to \eqref{chijump}, generically triggering a jump in the spacetime curvature. This means we have a landscape of possible vacua with different cosmological constants, scanned through membrane nucleation.

In order to be compatible with observations, this landscape must include vacua whose curvature is of order the Hubble scale today, $H_0^2$. This suggests a dense landscape with vacua separated by no more than this very low scale. In the original Brown-Teitelboim scenario \cite{Brown:1987dd,Brown:1988kg}, this required the membrane charge to be exponentially small in Planck units. That condition was lifted  in Bousso-Polchinski \cite{Bousso:2000xa} where a sufficiently dense landscape follows from a family of $\mathcal{O}(100)$ three-form fields coupled to membranes with incommensurate charges. In  Kaloper-Westphal \cite{Kaloper:2022oqv,Kaloper:2022jpv} the dense landscape follows from an irrational ratio for the fundamental membranes charges.

Of course, the density of the landscape raises the possibility that transitions between vacua of similar curvature are possible. If this is the case, we may encounter an empty universe problem where the universe  arrives at a low scale vacuum via a slow cascade through higher-curvature vacua, cooling the universe down through many many efolds of exponential expansion. Although the empty universe problem is known to be an issue for the original Brown-Teitelboim scenario, we will show how it is avoided in the other cases we consider. 

Membrane nucleation, necessary for scanning the landscape of vacua,  is a quantum process.  When we compute transition rates we do so between eigenstates of constant $\chi_i$. This suggests a path integral formalism equipped with Neumann boundary conditions, fixing $\chi_i$ in both  the in-state and in the out-state \cite{Duncan:1989ug}.  We will compute these transition rates in the next section. Corresponding formulae for Dirichlet boundary conditions, fixing $A^i$ in in and out states, are presented in the appendix.    

\section{Nucleation rates} \label{sec:rates}
To compute the rate at which membranes are nucleated, mediating transitions between vacua, we analytically continue to Euclidean signature, setting 
\be
t \to   -i t_E, \qquad S \to iS_E
\ee
where $S_E$ is the Euclidean action.  Following \cite{Brown:1987dd,Brown:1988kg}, we assume that scalars are unchanged under this Wick rotation, including the dual of the four-forms $\star F \to \star F$. It follows that the three-forms should be analytically continued as $A\to iA$ \footnote{If $\eta$ and $\eta_E=i\eta$ are the volume forms for Lorentzian and Euclidean signature respectively, then $F= -(\star F)\eta$ and $F_E=(\star F)_E \eta_E$.  Setting $\star F=(\star F)_E$, as desired, we infer that $F=iF_E$. This is consistent with our choice of $A=iA_E$.}. The resulting Euclidean action now takes the form
\begin{multline} \label{euclideanaction}
S_E =-\int_\mathcal{M} d^4 x_E\sqrt{|g|} \left[\frac{M_{pl}^2}{2}  R-\frac12 \omega^{ij}(\phi) \nabla_\mu \phi_i \nabla^\mu  \phi_j-V(\phi) \right] 
\\+\int_\mathcal{M} \left[ -\frac{1}{2}Z_{ij} (\phi)F^i\wedge \star F^j+\sigma_i(\phi) F^i \right]
 +S^E_\text{boundary}+S^E_\text{membranes}
\end{multline}
where the boundary terms are chosen to be consistent with Neumann boundary conditons on the three-form fields ($\lambda=1$). Since we will only be considering bounce configurations that transition between two vacua, we assume that there is a single membrane, $\Sigma$, which is charged under  $A^i$ for some particular choice of $i=i_*$.  The  membrane charges under $A^i$ are given by $Q_{i}=\delta_{i i_*} Q_{i_*}$  where $Q_{i_*}=\pm q_{i_*}$  and the membrane tension is $T=\tau_{i_*}$.  As a result, the  (Euclidean) membrane action is given by
 \be
 S^E_\text{membranes}=- Q_{i_*}\ \int_{\Sigma} \alpha_\Sigma^{i_*} + T \int_{\Sigma} d^3 \xi_E \sqrt{| \gamma_{\Sigma} |}\, .
 \ee
 We consider $O(4)$ symmetric Euclidean field configurations, with metric
\be
ds^2=dr^2+\rho(r)^2 d \Omega_3
\ee
where $d \Omega_3=h_{ij} d\xi^i d\xi^j$ is the metric on a unit 3-sphere, Euclidean three-form potentials 
\be
A^i=A^i(r) \sqrt{|h|} d^3 \xi
\ee
and scalars $\phi_i=\phi_i(r)$. The radial coordinate is assumed to run from $r_\text{min}$ to $r_\text{max}$ (to be determined) and the membrane lies at $r=0$.  We label the interior geometry, where $r_\text{min}\leq r<0$ , by $\mathcal{M}_-$ and the exterior, where $0<r<r_\text{max}$,  by $\mathcal{M}_+$.

With this ansatz, the field equations  away from the membrane simplify to the
 following 
\begin{eqnarray}
 3 M_{pl}^2 \left[\frac{1}{\rho^2} -\left(\frac{\rho'}{\rho}\right)^2 \right]&=&V-\frac12 \omega^{ij} \phi'_i \phi'_j+\frac12 Z_{ij} \frac{A^i{}' A^j{}'}{\rho^6} \qquad \\
  M_{pl}^2 \left[\frac{1}{\rho^2} -\left(\frac{\rho'}{\rho}\right)^2 -2\frac{\rho''}{\rho}\right] &=&V+\frac12 \omega^{ij} \phi'_i \phi'_j+\frac12 Z_{ij} \frac{A^i{}' A^j{}'}{\rho^6} \\
  \chi_i'&=&0 \\
  \frac{1}{\rho^3} \left(\rho^3 \omega^{kj} \phi_j' \right)'&=& V^{,k}-\frac{1}{2}Z_{ij}{}^{, k} \frac{A^i{}' A^j{}'}{\rho^6}+\frac{A^i{}'}{\rho^3}  \sigma_i{}^{,k} \qquad
\end{eqnarray}
where $\chi_i=\sigma_i-Z_{ij} \frac{A^j{}'}{\rho^3}$.

Assuming all scalars are locally  constant and satisfy the constraints listed from \eqref{con1} to \eqref{con3}, this system is solved by
\be
\rho=\frac{\sin(k(\epsilon r+r_0)}{k}, \qquad \epsilon=\pm 1
\ee
where the curvature $k^2$ matches the curvature of the Lorentzian solution \eqref{con1}, and three-form potentials
\be
A^i(r)=A^i(0)+c^i\int_0^r dr \rho(r)^3. \label{Asol}
\ee
Note that the solution for $\rho$ extends to $k^2 \leq 0$ by analytic continuation.  The parameter $r_0$ is an integration constant, setting the radius of the 3-sphere at the membrane. 

For $k^2>0$, the geometry is that of a section of a 4-sphere and it turns out  that $\rho$ is invariant under $\epsilon \to -\epsilon$, $r_0 \to \pi/k-r_0$. This allows us to set $\epsilon=+1$, WLOG, while also assuming $r_0 \in [0, \pi/k]$.  The poles of the 4-sphere are located at  $r_\text{min}=-r_0$ and $r_\text{max}=\pi/k-r_0$.

For $k^2=0$ and $k^2<0$, the geometry is that of a section of a four-dimensional Euclidean plane or hyperboloid respectively.  Since $\rho$ is assumed to be non-negative at the stack, we take $r_0\geq 0$. Note that we can no longer fix $\epsilon$ WLOG, and must consider each sign separately.  For $\epsilon=+1$, $r_\text{min}=-r_0$ corresponding to the point where the 3-spheres shrink to zero size, while $r_\text{max}=\infty$ corresponding to where they diverge.  For $\epsilon=-1$, the reverse is true: $r_\text{min}=-\infty$ corresponding to the point where the 3-spheres diverge, while $r_\text{max}=r_0$ corresponding to where they shrink to zero size. 

At the membrane, the induced metric $ds_{\Sigma}^2=\rho^2(0) d\Omega_3$ and the pullback of the three-forms $\alpha_\Sigma^i=A^i(0) \sqrt{|h|} d^3 \xi$ are well defined.  Note that for the induced metric this gives us a continuity constraint
\be
\Delta \rho(0) =\Delta \left[ \frac{\sin k r_0}{k} \right]=0\, ,
\ee
where we introduce the notation  $\Delta x=x^+-x^-$ and $\langle x\rangle=\frac12 (x^++x^-)$ corresponding respectively to the difference and average of some quantity $x$ defined on either side of the membrane.  We also have the following junction conditions on the conjugate momenta, 
\begin{eqnarray}
6M_{pl}^2 \Delta \left[ \frac{\rho'(0)}{\rho(0)}\right] &=&-3T \\
\Delta \chi_i &=&-\delta_{i i_*} Q_{i_*}. \label{Qjunc}
\end{eqnarray}
Physically realistic  membranes  always carry non-negative tension, resulting in the following constraint on the allowed configurations
\be
\Delta \left[ \epsilon \cos k r_0 \right] \leq 0\, . \label{Tcon}
\ee
The instanton solutions can also be classified according to the value of  
\be
X=\frac{4 M_{pl}^4 \Delta k^2}{T^2}
\ee
with $X>0$ in downward transitions and $X<0$ in upward transitions.  More importantly, however, it turns out that some configurations are kinematically allowed only when $|X|\leq 1$ with the remainder allowed when $|X|\geq 1$. An analogous result given explicitly in terms of membrane charges was shown to be true in  \cite{Kaloper:2022oqv,Kaloper:2022jpv}. Here we show how it is generalised to a much wider class of models which could,  in principle,  include the Bousso-Polchinski set-up \cite{Bousso:2000xa}. Indeed, we can always relate $X$ to the charge of the membrane and the curvature of the parent vacuum using equations \eqref{con1} to \eqref{con3}, along with the junction condition \eqref{Qjunc}, although the relationship is not always a simple one\footnote{In principle, this can be done as follows for a generic model. First, we use \eqref{con2} and \eqref{con3} to express the fluxes $c^i$ and the scalars $\phi_i$. in terms of $\chi_i$. Plugging this into \eqref{con1} we obtain an expression for the bulk cosmological constant of the form $k^2=\mathcal{K}(\chi_i)$. For example, for tunnelling from near Minkowski into AdS, we then have $\chi_i^+$ fixed such that $k^2_+=\mathcal{K}(\chi_i^+)\approx 0$, with $\chi_i^-=\chi_i^++Q_i$ fixed by the junction conditions \eqref{Qjunc}. Finally we have that $X^2=4 M_{pl}^4 \Delta k^2/T^2=4 M_{pl}^4 [\mathcal{K}(\chi_i^+)-\mathcal{K}(\chi_i^++Q_i)]/T^2$}.

To see how $X$ relates to the geometry of the instanton we use the junction conditions, 
\be
\Delta \rho(0)=0, \qquad \Delta [\rho'(0)]=-\frac{T}{2M_{pl}^2} \rho(0)  \label{deltarho1}
\ee
along with the expression $\rho'^2=1-k^2 \rho^2$ to prove the following useful relations
\be
\langle \rho'(0) \rangle = \frac{M_{pl}^2}{T} \rho(0)  \Delta k^2  \label{avrho1}
\ee
and
\be
\rho(0)=\frac{\frac{T}{M_{pl}^2}}{\sqrt{(\Delta k^2)^2+\frac{T^2}{M_{pl}^4} \langle k^2 \rangle +\left(\frac{T^2}{4M_{pl}^4}\right)^2}}. \label{rho0}
\ee
We now take the ratio of \eqref{deltarho1} and \eqref{avrho1} to see  that
\be
X=-\frac{2\langle \rho'(0) \rangle}{ \Delta [\rho'(0)]}=\frac{1+z}{1-z}, \qquad z=\frac{(\epsilon \cos kr_0)_+}{(\epsilon \cos kr_0)_-}.
\ee
The value of $X$ now corresponds to a kinematic constraint on the geometry. As an illustrative example, consider an instanton describing a $\text{dS}_+ \to \text{dS}_-$ transition with $(kr_0)_+ \geq \frac{\pi}{2} \geq (k r_0)_-$ and, of course, $\epsilon_\pm=1$. Since $(\cos kr_0)_+  \geq 0 \geq (\cos k r_0)_-$ it immediately follows that this solution has $T\geq 0$ and $z\leq 0$, or in other words, $-1 \leq X \leq 0$. 

A summary of these classifications is presented for all potential instanton geometries in table \ref{tab1}. The table also lists those geometries for which the corresponding tunnelling rates are  infinitely suppressed and so can be ignored. As we will see presently, these correspond to  configurations with $k_+^2 \leq 0$ and  $\epsilon_+=-1$. Table \ref{tab1} mirrors the `Baedeker' of  \cite{Kaloper:2022oqv}, although we emphasize again how it applies to a much broader set of models.
\begin{center}
\begin{table}[t]
\centering 
\resizebox{1.1\textwidth}{!}{%
\begin{tabular}{ |c||c|c|c|}
\hline
& $\text{dS}_+$& $\text{Minkowski/AdS}_+$ & $\text{Minkowski/AdS}_+$\\
& $ \epsilon_+=+1$&  $\epsilon_+=+1$ & $\epsilon_+=-1$\\
\hline
\hline
\begin{tabular}[t]{c}$\text{dS}_-$ \\  $\epsilon_-=+1$ \end{tabular}
& 
\begin{tabular}[t]{@{}c|c@{}}
         $(kr_0)_+ \geq \frac{\pi}{2} \geq (k r_0)_-$ & 
         $(kr_0)_+ \geq (k r_0)_-  \geq \frac{\pi}{2}$ \\
         allowed for $|X| \leq 1$  &
          allowed for $X \leq -1$
          \\ \hline
           $ \frac{\pi}{2} \geq (kr_0)_+ \geq (k r_0)_-$   & $(kr_0)_+ < (k r_0)_-$  \\ 
          allowed for $X \geq 1$  & 
         negative tension
      \end{tabular}
& negative tension &
 \begin{tabular}[t]{@{}c|c@{}}  
         $(kr_0)_- \geq \frac{\pi}{2}$ & 
         $\frac{\pi}{2} \geq  (kr_0)_-  $ \\
         kinematically & kinematically \\
         allowed for $X \leq -1$,  &
          allowed for $-1 \leq X \leq 0$, \\
           infinitely suppressed &  infinitely suppressed
            \end{tabular}
  \\
 \hline
 \begin{tabular}[t]{c}$\text{Minkowski/AdS}_-$\\ $\epsilon_-=+1$ \end{tabular}&  \begin{tabular}[t]{@{}c|c@{}}
         $(kr_0)_+ \geq \frac{\pi}{2}$ & 
         $\frac{\pi}{2} \geq  (kr_0)_+ $ \\
allowed for $0\leq X \leq 1$  &
~~allowed for $X \geq 1 $~~~
            \end{tabular} &
 \begin{tabular}[t]{@{}c|c@{}}  
         $|k_-| \geq |k_+|$ & 
         $|k_-| < |k_+|$ \\
         allowed for $X \geq 1$ &
         negative tension
            \end{tabular}
 &  \begin{tabular}[t]{c} kinematically allowed for $|X| \leq 1$,\\ infinitely suppressed \end{tabular} \\ \hline
\begin{tabular}[t]{c} $\text{Minkowski/AdS}_-$\\$\epsilon_-=-1$ \end{tabular} &  negative tension  & negative tension & \begin{tabular}[t]{@{}c|c@{}}  
         $|k_-| > |k_+|$ & 
         $|k_-| \leq |k_+|$ \\
         negative tension
          &  kinematically \\ & ~~allowed for $X \leq -1$, \\ &~~ infinitely suppressed
            \end{tabular}
 \\
 \hline
\end{tabular}%
}
\caption{Summary of transitions $\mathcal{M}_+ \to \mathcal{M}_-$ where $\mathcal{M}$ is de Sitter, Minkowski or anti de Sitter, as indicated. If a transition is forbidden because it violates the condition of non-negative tension, we mark it accordingly.  For the remaining transitions, we mark whether they are kinematically allowed for $|X| \leq 1$ or $|X|\geq 1$ where $X=4M_{pl}^4 \Delta k^2/T^2$.   In some of these cases, the transition may be kinematically allowed but is ruled out because the transition rate is infinitely suppressed.  Recall that we can take $\epsilon=+1$ for all dS configurations WLOG, so any examples contrary to this are not applicable. For a given model, equations \eqref{Qjunc} and \eqref{deltarho1}  can be used to determine, $X$ and identify precisely where we are on the Baedeker. } \label{tab1}
\end{table}
\end{center}

In semi-classical theory of vacuum decay, transition rates between vacua $\mathcal{M}_+ \to \mathcal{M}_-$ are given by \cite{Coleman:1977py, Callan:1977pt, Coleman:1980aw}
\be
\frac{\Gamma}{\text{Vol}} \sim e^{-B/\hbar}
\ee
where 
\be
B=S_E(\text{instanton})-S_E(\text{parent}).
\ee
Here $S_E(\text{instanton})$ is the Euclidean action evaluated on the bubble configurations described above, interpolating between the vacua $\mathcal{M}_+$ and $\mathcal{M}_-$.  In contrast, $S_E(\text{parent})$ is the Euclidean action evaluated on the complete parent vacuum,  $\mathcal{M}_+$, with no bubbles. 

The tunnelling exponent can be computed in all cases after a relatively lengthy calculation and plenty of heart warming cancellations,  giving 
\be
B=-2 M_{pl}^2 \Omega_3 \Delta \left\{\frac{1}{k^2} \left[ \rho'(r)^3 \right]^0_{r_\text{min}} \right\}+\Omega_3 T\rho(0)^3 \label{BN}
\ee
where  $r_\text{min}$ denotes the minimal value of the radial coordinate and  $\Omega_3$ is the volume of the unit 3-sphere. Note that this is quantitatively identical to the tunnelling exponent computed in General Relativity for the same geometry, reflecting the fact that the scalars are constants in vacuo, while the three-forms gravitate like a cosmological constant.  
Whenever we have a parent AdS  or Minkowski vacuum with $\epsilon_+=-1$, there are configurations with non-negative membrane tension whose tunnelling exponents contain divergent contributions at $r_\text{min}$.  These correspond to exotic configurations where an infinite asymptotic space is removed by membrane nucleation. The corresponding tunnelling rates are either infinitely suppressed, ruling out the transition, or infinitely enhanced, signalling a catastrophic instability. As we are imposed Neumann boundary conditions on the three-forms, these particular configurations are always infinitely suppressed\footnote{For $\text{AdS}_+ \to \text{AdS}_-$ configurations with $\epsilon_\pm=-1$, there are divergent contributions to the tunnelling exponent for both vacua.  If we assume that the divergence in the proper distance occurs at the same rate in each case, $|r_\text{min}^+| \sim |r_\text{min}^-| \to \infty$, then the positivity of the membrane tension ensures that $|k|_+>|k|_-$, and the dominance of the divergent contribution of the + vacuum.}.

As expressed in table \ref{tab1}, there are now only three configurations that avoid any problems with negative brane tension or infinitely suppressed tunnelling rates. These are the configurations of physical interest corresponding to
\begin{itemize}
    \item $\text{dS}_+ \to \text{dS}_-$
    \item $\text{dS}_+ \to \text{Minkowski/AdS}_-$
    \item $\text{Minkowski/AdS}_+ \to \text{Minkowski/AdS}_-$  $(|k_-| \geq |k_+|)$
\end{itemize}
 each with $\epsilon_\pm=1$ and so $\rho'(r_\text{min})=1$. To simplify the corresponding formulae for the tunnelling rates, we make use of equations \eqref{deltarho1} to \eqref{rho0}.  The tunnelling exponents for the three configurations of interest can now be written as
\begin{small}
\be
B=\frac{2M_{pl}^2 \Omega_3}{k_+^2 k_-^2}\left\{ -\Delta k^2  +\frac{M_{pl}^2}{T} \rho(0)\left[(\Delta k^2)^2+\frac{T^2}{2M_{pl}^4} \left \langle k^2 \right \rangle  \right]
   \right\} \label{BNphys}
\ee
   \end{small}
 If we recall that   $X=4M_{pl}^4 \Delta k^2/T^2$ and introduce 
 \be
 Y(X)=\sqrt{(X-1)^2+16 k_+^2 M_{pl}^4/T^2}
 \label{eq:defY}
 \ee
 we can rewrite the tunnelling exponent in the following simple form
\be
B=\frac{4M_{pl}^2 \Omega_3}{k_+^2 } \left[\frac{1+Y-X}{Y(1+Y+X)} \right]\, .
\ee
\section{Cascades, bungees and the stability of the Minkowski vacuum} \label{analysis}
Armed with a general expression for the tunnelling exponent and a detailed understanding of the instanton geometries as shown in table \ref{tab1}, we are ready to take a more in-depth look at the tunnelling dynamics. In particular, we are interested in what happens if the universe begins in a high scale de Sitter vacuum   with curvatures far above the current Hubble scale. Can we descend into the current low scale vacuum in a single jump, avoiding the empty universe problem, or do we need to cascade into it via a series of incremental changes in curvature? Is there a danger we fall deep into an anti-de Sitter vacuum, so that the universe is driven to a apocalyptic crunch?  How long is the current vacuum expected to survive before it gives way to these apocalyptic AdS vacua?

Consider tunnelling from a parent vacuum with a non-negative curvature, so that $k_+^2 \geq 0$. It follows that  $Y\ge |X-1| \ge 0 $, and so rewriting the tunnelling exponent as
\be
B=\frac{4M_{pl}^2 \Omega_3}{k_+^2 } \left[\frac{Y-(X-1)}{Y(2+Y+X-1)} \right], 
\ee
we see that it is everywhere positive\footnote{This contradicts \cite{Kaloper:2022oqv} where it is claimed that the bounce action can be negative when the membrane radius is close to the de Sitter radius of the two vacua. This involved neglecting terms that we said to be small compared to a ``leading" term, equal to the difference in curvature. However, this ``leading" term is just as small as the neglected terms in the relevant limit.}. Further, for $k_+$ fixed, it is monotonically decreasing from $4M_{pl}^2 \Omega_3/k_+^2$ as $X \to -\infty$ to zero as $X \to +\infty$\footnote{The quickest way to see this is to define $\zeta \in (-\infty, \infty)$ such that $X\equiv \frac{4M_{pl}^4 \Delta k^2}{T^2}=1+\frac{4M_{pl}^2 k_+}{T} \sinh \zeta$. The tunnelling exponent then reduces to the following simple form
$$
B=\frac{4M_{pl}^2 \Omega_3}{k_+^2 (1+e^{2\zeta})\left(1+\frac{2M_{pl}^2 k_+}{T} e^\zeta\right)}~.
$$ This is clearly monotonically decreasing between the two limits stated in the text. }. For a given parent de Sitter vacuum, this means that transitions towards smaller values of the curvature are likely to be the most rapid, provided they are kinematically allowed.  This raises the possibility of decay deep into anti de Sitter space, which would be incompatible with the current low scale vacuum, with curvature no larger than the current Hubble scale.

Generically we find that descent into AdS is kinematically forbidden for parent vacua with very large de Sitter curvatures. This is because the membrane  doesn't carry enough charge to trigger a sufficiently large jump in the cosmological constant.  For parent vacua with curvature less than some critical value, $k_+^2 \leq k_c^2$, descent into AdS is kinematically allowed.  This is potentially dangerous if the tunnelling rates are too high.  One way to avoid this is as follows: first we  choose the parameters of the model such that $0<X<1$ for all downward transitions from parent vacua below the critical curvature,  $k_+^2 \leq k_c^2$. It follows that
\be
 B > B(X=1)= \frac{2M_{pl}^2 \Omega_3}{k_+^2 \left(1+\frac{2M_{pl}^2 k_+}{T} \right)  }> \frac{M_{pl}^2 \Omega_3}{k_c^2}=B_c \label{Bc}
\ee
where we have also used the fact that   $4 M_{pl}^4 k_+^2/T^2<1+4 M_{pl}^4 k_-^2/T^2\leq  1$ for  $k_+^2 \leq k_c^2$ and $k_-^2 \leq 0$. 
Since we typically expect $k_c \lesssim M_{pl}$, we can choose parameters such that $B_c$ is large, and all vacua in danger of decay into anti de Sitter are long lived.  Furthermore, the lifetime of these vacua increases as we reduce the curvature of the parent vacuum, with $B$ containing a pole as $k_+ \to 0$.  

 We can see the importance of the proposed bound on $X$ by taking a closer look at transitions from parent vacua near Minkowski space. For these transitions, we have
\be
X_{M_+ \to AdS_-}\approx -\frac{4M_{pl}^4 k_-^2}{T^2}.
\ee
Note that $X_{M_+ \to AdS_-}=|X_{M_+ \to AdS_-}|$ since the daughter vacuum is AdS and so $k_-^2<0$. The tunnelling exponent for the transition goes as
\begin{equation}
B_{M_+ \to AdS_-} \sim \frac{2 M_{pl}^2 \Omega_3}{k_+^2}(1-S(X))+\frac{8 M_{pl}^6 \Omega_3}{T^2 X(X-1)^2}\left[(X-1)^2(1-S(X))+2S(X)\right]
\end{equation}
where $S(X)=\text{sgn}(X-1)$ and we understand that $X=X_{M_+ \to AdS_-}$ here and in the remainder of this section.
 The lifetime of the parent near Minkowski vacuum is clearly determined by the sign of $X-1$. In particular, when $X>1$, we have 
\be
B_{M_+ \to AdS_-}  \sim \frac{16 M_{pl}^6 \Omega_3}{T^2 X(X-1)^2}, \qquad X>1.
\ee
For tunnelling into deep AdS so that $X$ is large, this exponent is suppressed and the transition is quick, rendering the near Minkowski vacuum short lived. In contrast, if $X<1$, the tunnelling exponent goes as  
\be
B_{M_+ \to AdS_-} \sim \frac{4 M_{pl}^2 \Omega_3}{k_+^2},\qquad X<1.
\ee
As the parent vacuum approaches Minkowski, this exponent diverges, suppressing any transition into AdS and ensuring a long lived Minkowski vacuum. This is the key ingredient for addressing the cosmological constant advocated in \cite{Kaloper:2022oqv,Kaloper:2022jpv}, although we now see how it can apply more generally, as long as the microscopic details of transition from Minkowski into AdS enforce $X_{M_+ \to AdS_-}=-\frac{4M_{pl}^2 k_-^2}{T^2}<1$. 

Of course, it is not enough for the near Minkowski vacuum to be long lived. We need to show that we are likely to reach it from a generic initial state corresponding to a high scale de Sitter vacuum. There are two ways in which this might happen. The first involves a bungee jump, a single transition from a high scale de Sitter vacuum towards a near Minkowski vacuum, $k_+^2 \gg  H_0^2 \gtrsim |k_-|^2$. If a bungee jump is kinematically possible, the tunnelling exponent goes as
\be
B_{dS_+ \to M_- } \sim \frac{4 M_{pl}^2 \Omega_3}{k_+^2\left(1+\frac{4 M_{pl}^4 k_+^2}{T^2}\right)^{2}}.  \label{largejump}
\ee
For de Sitter vacua, the larger the value of $k_+$, the more this exponent is suppressed and the faster the transition.  Bungee jumps avoid the empty universe problem \cite{Abbott:1984qf}.  As explained in \cite{Bousso:2000xa}, the inflaton is significantly displaced from its minimum by quantum diffusion before the transition. After the transition, the inflaton is in slow roll and follows the classical evolution towards the minimum, at which point it begins to oscillate and reheating occurs. However, because the vacuum energy is now no larger than the current Hubble scale, there is no rapid cooling. 

The second way in which we might finally arrive at the current vacuum is via a cascade, several transitions in which the cosmological constant descends from a high scale in small increments, each with $\Delta k^2 \lesssim H_0^2$. For small incremental transitions where the vacuum energy changes by a small amount, $k_+^2 \to k_-^2=k_+^2(1-\epsilon)$, for $\epsilon \ll 1$,  the tunnelling exponent goes as
\be
B_{dS_+ \to dS_- \approx dS_+} \sim \frac{4 M_{pl}^2 \Omega_3}{k_+^2\sqrt{1+\frac{16 M_{pl}^4 k_+^2}{T^2}}}.   \qquad \label{smalljump}
\ee
Unfortunately, these cascades do run into the empty universe problem  \cite{Abbott:1984qf}.  As the cosmological constant slowly descends, quantum diffusion effects are scaled down, allowing the inflaton to settle into its minimum and  reheat the universe. However, this happens too soon, before the vacuum energy has settled into its current value. The universe then spends far too long in de Sitter vacua at scales above the current Hubble scale, and any matter produced after reheating is diluted away by exponential expansion. 

To avoid the empty universe problem,  it is clear that bungee jumps with $k_+^2 \gg  H_0^2 \gtrsim |k_-|^2$  must be kinematically allowed.  Furthermore, a cascade of incremental transitions with $\Delta k^2 \lesssim  H_0^2$, should either be kinematically forbidden, or suppressed relative to a bungee jump. If this is the case, the universe  initially in a high scale de Sitter vacuum is expected to find the current vacuum before inflation has ended and reheating has begun. Furthermore, if we can consistently impose a condition $|X| <1$ for a parent vacuum close to Minkowski, then the current vacuum is long lived, and the catastrophe of descent into AdS is delayed until the far future, or avoided altogether.
\section{Examples} \label{sec:examples}
As we have seen, the key features for addressing the cosmological constant problem proposed in \cite{Kaloper:2022oqv,Kaloper:2022jpv} could apply quite generally to a large class of models under consideration. This is because the scalars and the four-form field strengths all gravitate as cosmological constants in the vacuum. Consequently, the tunnelling exponent, $B$, takes a universal form which we can express in terms of the tension of the membrane and the curvature of the two vacua involved in the transition. Things only become model dependent when we ask how the quantities relate to other model parameters, such as the membrane charge. In particular,  the condition $|X|<1$ will mean different things for different models. Kinematic considerations can also vary. We shall now explore a few cases of particular interest. 
\subsection{The Brown-Teitelboim set-up}
We begin with the pioneering model of this class, originally proposed by Brown and Teitelboim \cite{Brown:1987dd,Brown:1988kg}. As is well known, this model runs into problems with the empty universe and cannot be taken seriously as a solution to the cosmological constant problem.  However, we include a discussion of its main features as a warm up to the more interesting models.  In the Brown-Teitelboim set-up,  there are no scalars and a single four-form, corresponding to  the case where $i, j=1$ only and 
\be
\omega^{ij}=\sigma_i=0,\quad Z_{ij}=1, \quad V=V_{QFT}
\ee
where $V_{QFT}$ is the renormalised vacuum energy.  In the absence of scalars, the bulk constraints \eqref{con1} to \eqref{con3} yield 
\begin{eqnarray}
&& 3M_{pl}^2 k^2 = V_{QFT}+\frac12 c^2   \label{con1BT} \\
&&\chi = -c \label{con3BT}
\end{eqnarray}
from which it follows that
\be
k^2=\frac{V_{QFT} +\frac12 \chi^2}{3M_{pl}^2}\ .
\ee
Membrane nucleation triggers a jump $\Delta \chi=-Q=-\eta q$, where $q$ is the fundamental membrane charge and $\eta=+1$ for branes and $\eta=-1$ for anti-branes. This is consistent with the quantisation condition $\chi=-Nq$ for $N \in \mathbb{Z}$, so that membrane nucleation gives $\Delta N=\eta$. The spectra of vacua are now characterised by  the following curvature
\be
k^2(N)=\frac{V_{QFT} +\frac12 N^2 q^2}{3M_{pl}^2}\ .
\ee
For this to include near Minkowski vacua, we assume $V_{QFT}<0$, which is easy to engineer in string compactifications. The smallest possible separation in curvature, $\delta k^2$, occurs between   neighbouring vacua and takes the value
\be
\delta k^2=k^2(N \pm 1)-k^2(N)=\frac{(1 \pm 2N)q^2}{6 M_{pl}^2}\ .
\ee
This density varies with where we are in the landscape, through the dependence on $N$.  For a near Minkowski vacuum, $N=N^\text{Mink} \approx  \pm \sqrt{2|V_{QFT}|}/q$ and the separation takes two possible values
\be
\delta k^2|_{k^2=0} \approx \frac{q^2 \pm q \sqrt{8 |V_{QFT}|} }{6 M_{pl}^2}\, .
\ee
To be compatible with observation, we require the absolute value of both to be no larger than the current Hubble scale, $H_0^2$.  Assuming  $|V_{QFT}| \gg M_{pl}^2 H_0^2$, this imposes an upper bound on the membrane charge
\be
q \lesssim  \frac{6 M_{pl}^2 H_0^2}{\sqrt{8 |V_{QFT}|} }\ . \label{qcon}
\ee
Nucleation of a membrane triggers a jump $N^+\to N^-=N^+ \pm 1$, where we take the $+$ for a brane and the $-$ for an anti-brane. In the Brown-Teitelboim setup, this corresponds to a minimal change in curvature
\be
\Delta k^2 =\frac{(1\mp 2N^-)q^2}{6 M_{pl}^2}=\frac{(-1\mp 2N^+)q^2}{6 M_{pl}^2}\ .
\ee
In a general set-up, adjacent vacua in flux space are separated by a curvature, $\Delta k^2$, which is generically {\it not} the same  as the smallest jump in curvature, $\delta k^2$. Crucially, however, in the Brown-Teitelboim set-up, the simplicity of the  model renders the two to be equivalent: $\Delta k^2=\delta k^2$.  By the density assumption, nearest neighbour jumps  are now  no larger than the current Hubble scale whenever the transitions include the current vacuum.  More explicitly, consider the transition from a parent de Sitter vacuum to a near Minkowski vacuum.  For this to the arise from the nucleation of a single membrane, the parent vacuum will have curvature
\be
k_+^2 =\frac{(1+2|N^\text{Mink}|)q^2}{6 M_{pl}^2}\approx  \frac{q}{6M_{pl} ^2} \sqrt{8 |V_{QFT}|} \lesssim H_0^2
\ee
where we have used the fact that  $|N^\text{Mink}| \gtrsim |V_{QFT}|/M_{pl}^2 H_0^2 \gg 1$ along with the density constraint
\eqref{qcon} on the fundamental charge.  This demonstrates the fact that bungee jumps to the current vacuum cannot be  mediated by single membrane.   As is well known, this suggests that the Brown-Teitelboim set-up suffers from an empty universe problem  \cite{Bousso:2000xa}. Tunnelling rates in this set-up are easily computed and can be shown to agree with \cite{Brown:1987dd,Brown:1988kg}. This can also be obtained from the corresponding analysis in the next section by taking the limit of a single three-form potential.

\subsection{The Bousso-Polchinski set-up}
We now switch to a set-up with no scalars and a large number of four-forms as originally proposed by Bousso and Polchinski \cite{Bousso:2000xa} as a way to overcome the limitations of \cite{Brown:1987dd,Brown:1988kg}. This corresponds to the case where
\be
\omega^{ij}=\sigma_i=0,\quad Z_{ij}=\delta_{ij}, \quad V=V_{QFT}
\ee
with $i, j$ running from $1$ to $\N$. In the absence of scalars, the bulk constraints \eqref{con1} to \eqref{con3} yield 
\begin{eqnarray}
&& 3M_{pl}^2 k^2 = V_{QFT}+\frac12 \delta_{ij}c^i c^j   \label{con1BP} \\
&&\chi_i = -\delta_{ij}c^j \label{con3KP}
\end{eqnarray}
from which it follows that
\be
k^2=\frac{V_{QFT} +\frac12 \delta^{ij} \chi_i \chi_j}{3M_{pl}^2}\ .
\ee
Recall the quantisation condition $\chi_i=-N_i q_i$ (no sum), for some $N_i \in \mathbb{Z}$. Assuming nucleation of a brane or anti-brane of type $i_*$, this is consistent with the membrane junction condition \eqref{Qjunc}, $\Delta \chi_i=-\delta_{i i_*} Q_{i_*}$, with $Q_{i_*} =\pm q_{i_*}$ (no sum), or equivalently, $\Delta N_i=\pm \delta_{ii_*}$.  The spectra of vacua is therefore characterised by curvature
\be
k^2 =\frac{V_{QFT} +\frac12 \sum_{i=1}^{\N} N_i^2 q_i^2}{3M_{pl}^2}\ .
\ee
For this to include near Minkowski vacua, we must once again assume that $V_{QFT}<0$. The vacua span an $\N$ dimensional grid with spacing $q_i$. Surfaces of constant vacuum curvature correspond to spheres centred at $N_i=0$. Consider two such surfaces of curvature $k^2= \frac{V_{QFT} +\frac12 r^2}{3M_{pl}^2} $ and $k'^2 =\frac{V_{QFT} +\frac12 r'^2}{3M_{pl}^2} $.  The corresponding spheres form the boundary of a shell of volume $V_\text{shell}\approx \frac12 \Omega_{\N-1} r^{\N-2}\delta r^2$ where $\delta r^2=r'^2-r^2$ and $\Omega_{\N-1}=2 \pi^{\N/2}/\Gamma(\N/2) $ is the volume of a unit $(\N-1)$-sphere. To estimate the density of vacua, we require that this shell should contain at least one grid point (or if there is degeneracy $D$, it should contain $D$ grid points), setting a bound $V_\text{shell} \gtrsim D \prod_{i=1}^\N q_i$. This now translates into vacua with a typical curvature separation of \cite{Bousso:2000xa}
\be
\delta k^2 \approx \frac{D \prod_{i=1}^\N q_i}{3 M_{pl}^2\Omega_{\N-1} r^{\N-2}}, \qquad r=\sqrt{6M_{pl}^2 k^2-2V_{QFT}}\ .
\ee
For this model, we assume the $q_i$ are incommensurate so that $D \sim 2^\N$, arising from the vacuum degeneracy $N_i \to -N_i$. As in the Brown-Teitelboim set-up, the density of vacuum energies depends on where you are in the landscape. Near Minkowski, the density is 
\be
\delta k^2|_{k^2=0} \approx \frac{2^\N \prod_{i=1}^\N q_i}{3 M_{pl}^2\Omega_{\N-1} (2|V_{QFT}|)^{\N/2-1}} \ .
\ee 
If we assume that there are no hierarchies in the membrane charges,  $q_i \sim \mathcal{O}(q)$ for all $i$, whilst remaining incommensurate, it turns out that 
\be
\delta k^2|_{k^2=0} \sim \frac{2 |V_{QFT}|}{3 M_{pl}^2}\sqrt{\frac{\pi}{\N}} \left( \frac{\N q^2}{e \pi |V_{QFT}| }\right)^{\N/2} \label{densityBP}
\ee 
Provided $\N q^2 < e \pi |V_{QFT}|$, by taking $\N$ sufficently large, we can achieve a density well within the desired observational bound $\delta k^2|_{k^2=0} \lesssim H_0^2$ where $H_0 \sim 10^{-60} M_{pl}$ is the current Hubble scale. For example, if $|V_{QFT}|\sim M_{pl}^4$ and $q\sim 0.02M_{pl}^2$ then we can achieve the desired density of vacuum energies for $\N \approx 100$.

Nucleation of a membrane of type $i_*$ triggers a jump in one of the flux integers, $N^+_{i_*} \to N_{i_*}^-=N^+_{i_*} \pm 1$, while leaving the remaining $N_{i \neq i_*}$ unchanged. This translates into a jump in vacuum curvature
\be
\Delta k^2=\frac{(1\mp 2N_{i_*}^-)q_{i_*}^2}{6 M_{pl}^2}=\frac{(-1\mp 2N_{i_*}^+)q_{i_*}^2}{6 M_{pl}^2}.
\ee
As in the previous example, this is sensitive to where we are in the landscape through the dependence on $N^\pm_{i_*}$, a consequence of the curvature having quadratic four-form dependence. 

Let us consider descent from a parent de Sitter vacuum with curvature $k_+^2 \geq 0$ and compute the important quantity, $X=4M_{pl}^4 \Delta k^2/T^2$.  To do this, we first relate the curvature to the integer fluxes using the relation
\be
\sum_{i=1}^\N (N_i^+)^2 q_i^2=r_+^2, \quad r_+=\sqrt{6 M_{pl}^2 k_+^2-2V_{QFT}}. \label{Nr}
\ee
Since $\Delta k^2 \geq 0$, it now follows that 
\be
X=\frac{2 M_{pl}^2 q_{i_*}^2}{3 \tau_{i_*}^2}(2|N_{i_*}^+|-1) \sim \frac{2 M_{pl}^2 q^2}{3 \tau^2}(2|N_{i_*}^+|-1)
\ee
where we have used the fact  there are no hierarchies, $q_i \sim \mathcal{O}(q)$, $\tau_i \sim \mathcal{O}(\tau)$ for all $i$. Further, equation \eqref{Nr} now implies that $|\vec N|^2 \sim r^2_+/q^2$, where $|\vec N|^2=\sum_{i=1}^\N N_i^2 \geq |N_{i_*}|^2$ and so
\be
X \lesssim \frac{4 M_{pl}^2 qr_+}{3 \tau^2}\equiv X_\text{max}(k_+)\, .
\ee
 Dangerous transitions into AdS are kinematically forbidden if $X_\text{max}(k_+)<X_0(k_+)$, where $X_0(k_+)=4 M_{pl}^4 k_+^2/\tau^2$ represents the critical value of $X$ for which the daughter vacuum is approximately Minkowski.  
For the largest values of $k_+^2 \gg  |V_{QFT}|/3 M_{pl}^2$, it is clear that $r_+ \approx \sqrt{6} M_{pl} k_+$ and so  
\be
X_\text{max} \approx \frac{4\sqrt{6}  M_{pl}^3 qk_+}{3 \tau^2} < \frac{4 M_{pl}^2  k_+}{3 \tau^2} \sqrt{\frac{6 e \pi |V_{QFT}|}{\N}}\ll X_0(k_+)
\ee
where we have used the density condition $ \N q^2 < e \pi |V_{QFT}|$. It is, therefore, kinematically impossible for parent vacua with $k_+^2 \gg  |V_{QFT}|/3 M_{pl}^2$ to decay directly into AdS.  

What about parent vacua with $k_+^2 \lesssim  |V_{QFT}|/3 M_{pl}^2$? These  have $r_+ \sim \sqrt{2|V_{QFT}|}$ and so
\be
X_\text{max} \sim \frac{2  M_{pl}^2 q  }{3 \tau^2} \sqrt{8|V_{QFT}|}.\label{XmaxBP}
\ee
For sufficiently small $k_+^2<k_c^2=q\sqrt{8|V_{QFT}|}/ 6M_{pl}^2 $, we can have $X_\text{max}> X_0(k_+)$ and  so decay into AdS is now kinematically possible.  

The potential for decay into AdS might be troubling at first glance. However, as we hinted in the previous section, we can impose the condition
\be
\frac{2 M_{pl}^2   q} {3\tau^2}  {\sqrt{8 |V_{QFT}|}}<1\ \label{condBP}
\ee
ensuring that $0<X<1$ whenever the parent vacuum has $k^2_+<k_c^2$ and is therefore vulnerable to this dangerous decay. However, the lifetime of these vacua can be made large by ensuring that $B_c \gg 1$, where $B_c$ is defined by equation \eqref{Bc}. Inputting the value of $k_c$ we have just derived for the Bousso-Polchinski set-up, we find that
\be
B_c \gtrsim \frac{6M_{pl}^4 \Omega_3}{q \sqrt{8 |V_{QFT}|}    }
\ee 
For our canonical example, with $|V_{QFT}|\sim M_{pl}^4$, $q\sim 0.02M_{pl}^2$  and  $\mathcal{N} \approx 100$, we have $B_c \gtrsim 2000$, and so vacuum decay into AdS is heavily suppressed in all instances.  Furthermore, the condition \eqref{condBP} ensures that $|X_{M_+ \to AdS_-}|<1$, guaranteeing the presence of a pole in the transition rate as the parent vacuum approaches Minkowski, rendering it arbitrarily long-lived. 

All of this suggests that the current vacuum can be selected using probabilistic arguments, similar to those presented  in \cite{Kaloper:2022oqv,Kaloper:2022jpv}. This is a significant departure from the anthropic arguments  usually presented for the Bousso-Polchinski set-up  \cite{Bousso:2000xa}. There is a small price to pay for this success: the condition \eqref{condBP} is in violation of the weak gravity conjecture for membranes \cite{Ibanez:2015fcv}, which requires 
\be
\tau_i\lesssim g_i M_{pl} q_i \label{wgc}
\ee
where $g_i$ is the strength of the three-form coupling. To see this, note that we have two inequalities on the charge, 
\be
\sqrt{\frac{e\pi |V_{QFT}|}{\mathcal{N}}}>q >\frac{2 M_{pl}^2   q^2}{3\tau^2}  \sqrt{8 |V_{QFT}|}
\ee
and so 
\be
\frac{M_{pl} q}{\tau}<\left(\frac{9 e \pi}{32\mathcal{N}}\right)^{1/4}< 1
\label{eq:BPWGC}
\ee
in violation of \eqref{wgc}. 

If we assume sub-Planckian membrane tension consistent with a well-defined effective theory and the maximum charge, $q_\text{max}$,  tolerated by the density constraint $\delta k^2 \lesssim H_0^2$ (where $\delta k^2$ is given by equation \eqref{densityBP}), the stability condition \eqref{condBP} is also incompatible with an arbitrarily large  underlying vacuum energy $|V_{QFT}|$.   To see this, we note that 
\be
q_\text{max}=\sqrt{\frac{e \pi |V_{QFT}|}{\mathcal{N}}} \left(\frac{3 M_{pl}^2 H_0^2}{2 |V_{QFT}| }\sqrt{\frac{\mathcal{N}}{\pi}} \right)^\frac{1}{\mathcal{N}} \approx  \sqrt{\frac{e \pi |V_{QFT}|}{\mathcal{N}}} \epsilon (\N), \quad 
\ee
where,  in the last step, we have taken a limit of large $\N$ and defined  $\epsilon(\N) =\left( \frac{H_0^2}{M_{pl}^2}\right)^\frac{1}{\N}<1 $. We include the $\epsilon(\N)$ correction to the large $\N$ limit to allow for the fact there is a large hierarchy between $H_0$ and $M_{pl}$. For $q \sim q_\text{max}$, the stability condition \eqref{condBP} now implies that
\be
\frac{2|V_{QFT}|}{3 M_{pl}^4} \sqrt{\frac{8 e \pi}{\mathcal{N}}}\left(\frac{3 M_{pl}^2 H_0^2}{2 |V_{QFT}| }\sqrt{\frac{\mathcal{N}}{\pi}} \right)^\frac{1}{\mathcal{N}} \lesssim  \frac{\tau^2}{M_{pl}^6}<1
\ee
where we have also used the condition of sub-Planckian membrane tensions.  This can be used to place a limit on the scale of the QFT contribution to the cosmological constant. For large $\N$, this bound on $|V_{QFT}|$ takes on a particularly simple form 
\be
|V_{QFT}| \lesssim \frac{3}{2\epsilon(\N)} \sqrt{\frac{\N}{8 e \pi}} \times  \frac{\tau^2}{M_{pl}^2}.
\ee
This is quadratically sensitive to the brane tension and not especially restrictive as can be seen from the numerical examples in table \ref{tab:BP}. Note, in particular,  the case where  $\N=100$, corresponding to the benchmark point in \cite{Bousso:2000xa}. Here we can have Planckian $|V_{QFT}|$ with the current vacuum being both natural and quantum mechanically stable if the brane tensions, $\tau \sim 0.2   M_{pl}^3$ and the charges, $q\le 0.02  M_{pl}^2$.
\begin{table}[h!]
\begin{center}
    \begin{tabular}{|c|c|c|}
    \hline 
        $\mathcal{N}$ & Upper bound on   $|V_{QFT}|$ & Upper bound on $q_\text{max}$  \\
        \hline \hline
         10 &  $10^{12} \times  \frac{\tau^2}{M_{pl}^2}$ & $10^{-7} \times \frac{\tau}{M_{pl}}$ \\ \hline 
         60 &  $10^{2} \times  \frac{\tau^2}{M_{pl}^2}$ & $10^{-2} \times \frac{\tau}{M_{pl}} $ \\ \hline 
         100 & $10 \times  \frac{\tau^2}{M_{pl}^2}$ & $10^{-1}  \times \frac{\tau}{M_{pl}}$ \\
             \hline
    \end{tabular}
    \caption{Maximum scale of QFT contribution to the cosmological constnat (in units of $\tau^2/M_{pl}^2$) compatible with the quantum mechanical stability of low scale de Sitter space. We also show the  corresponding upper bound on the brane charges (in units of $\tau/M_{pl}$). The quoted numbers are order of magnitude estimates.}\label{tab:BP}
    \end{center}
\end{table}

The last thing we need to check is already well known: is it possible to transition from a high scale de Sitter vacuum to  the current vacuum in a single jump?  Near Minkowski daughter vacua satisfy 
\be
\sum_{i=1}^\mathcal{N} (N_i^\text{Mink})^2 q_i^2 \approx 2|V_{QFT}|.
\ee
If this is the daughter vacuum generated by the nucleation of a brane of type $i_*$, then the parent vacuum has curvature
\be
k_+^2=\frac{\sum_{i=1}^\mathcal{N} (1+ 2 |N_{i_*}^\text{Mink}|)q_{i_*}^2 }{6M_{pl}^2}   \sim \frac{q\sqrt{8 |V_{QFT}| }}{6 M_{pl}^2}=k_c^2
\ee 
For physically allowed configurations with positive brane tension this will be positive, corresponding to  a {\it high} scale de Sitter vacuum provided $q$ is not tuned to exponentially small values. In particular, for our canonical parameter values, $|V_{QFT}|\sim M_{pl}^4$, $q\sim 0.02M_{pl}^2$  and  $\mathcal{N} \approx 100$, we have a bungee jump from vacua of curvature $k_+^2 \sim 0.01 M_{pl}^2$ to the current vacuum. This enables us to avoid the empty universe problem in the Bousso-Polchinski set-up, as is well known. 

To summarise, if we are prepared to accept some violation of the membrane weak gravity conjecture, we can select the near Minkowski vacuum using probabilistic methods even for the Bousso-Polchinski set-up. This offers a tantalising alternative to the standard anthropic arguments used to recover a vacuum consistent with observation.  

\subsection{The Kaloper-Westphal set-up}
We now switch to the original motivation for our work: the Kaloper-Westphal set-up \cite{Kaloper:2022jpv}. This corresponds to the case where there are just two three-forms (so $i,j$ runs from $1$ to $2$) and 
\be
\omega^{ij}= Z_{ij}=0, \quad V=V_{QFT}+M_{pl}^2(\phi_1+\phi_2), \quad \sigma_i=-2\phi_i.
\ee
Crucially, the membrane charges are assumed to have an irrational ratio, $q_1/q_2=w \notin \mathbb{Q}$. The bulk constraints on the vacua \eqref{con1} to \eqref{con3} now give
\begin{eqnarray}
&& 3M_{pl}^2 k^2 = V_{QFT}+M_{pl}^2 (\phi_1+\phi_2)  \label{con1KW} \\
&& M_{pl}^2=2 c^k\label{con2KW}\\
&&\chi_i = -2\phi_i \label{con3KW}
\end{eqnarray}
from which it follows that
\be
k^2=\frac{V_{QFT}-\frac{M_{pl}^2}{2}(\chi_1+\chi_2)}{3M_{pl}^2}.
\ee
Recall that the nucleation of the membrane of type $i_*$ triggers a jump $\Delta \chi_i=-\delta_{ii_*} Q_{i_*}$, where $Q_{i_*}=\pm q_{i_*}$, consistent with the quantisation condition on $\chi_i=-N_i q_i$ (no sum) for some $N_i \in \mathbb{Z}$. The spectra of vacua is characterised by the curvature 
\be
k^2=\frac{V_{QFT}+\frac{M_{pl}^2}{2}(N_1 q_1+N_2 q_2)}{3M_{pl}^2}.
\ee
The irrational ratio in membrane charge ensures that this landscape of vacua is dense and passes arbitrarily close to the Minkowski vacuum. The nucleation of our membrane triggers a consistent jump $N^+_{i_*} \to N_{i_*}^-=N^+_{i_*} \pm 1$ leaving $N_{i \neq i_*}$ unchanged.  This gives rise to  a jump in the vacuum curvature
\be
\Delta k^2=\pm \frac{ q_{i_*}}{6}
\ee
giving
\be
X=\pm \frac{2 M_{pl}^4 q_{i_*}}{3 \tau^2_{i_*}}.
\ee
If we start in a parent de Sitter vacuum with curvature $k_+^2$, transitions to AdS are only possible for $k_+^2<k_c^2=q/6$, where $q=\text{max}\{q_{1}, q_2\}$. In \cite{Kaloper:2022jpv}, the authors impose the condition, 
\be
\frac{2 M_{pl}^4 q_i}{3 \tau_i^2}<1, \qquad i=1,2 \label{condKW}
\ee
which is equivalent to  $|X|<1$ in all cases. In particular, as we have seen in the previous section, this type of condition ensures that those vacua vulnerable to decay into anti de Sitter can be long lived. More precisely, when $k_+^2 < k_c^2=q/6$, we recall that the tunnelling exponent $B \geq B_c$, where $B_c$ is computed using \eqref{Bc}, 
\be
B_c\sim \frac{6 M_{pl}^2 \Omega_3}{q}.
\ee
As long $q$ is sub-Planckian, this remains large. Indeed, for $q_\text{max} \sim 0.1 M_{pl}^2$, the low scale de Sitter  vacua can be made very long lived, with $B_c \gtrsim 1000$. Since $|X|<1$, we also have  absolute stability of the Minkowski vacuum thanks to the presence of the pole in the tunnelling exponent as $k_+ \to 0$.  Note that unlike in the Bousso-Polchinski setup, the condition  \eqref{condKW}  is independent of $|V_{QFT}|$ or, in other words, the quantum mechanical stability of the Minkowski vacuum can be guaranteed whatever the depth of the bare AdS vacuum. 

The Kaloper-Westphal model also avoids the empty universe problem. This is because we  can bungee jump to the current vacuum from a parent de Sitter vacuum with curvature $k_+^2 \sim q/6$, where $q$ is not constrained to be exponentially small in Planck units.  Assuming sub-Planckian brane tension,  we might worry that the one condition we do have \eqref{condKW} is at odds with the weak gravity conjecture for membranes \eqref{wgc}.   However, in the absence of kinetic terms for the three-forms in the Kaloper-Westphal set-up, it is not immediately clear how one should implement the weak gravity condition in the first place. Of course, the absence of the kinetic terms would itself present a challenge to string theory model builders.

\subsection{More fun with irrationals}
The key ingredients of the Kaloper-Westphal set-up can be realised in a simple model of a single axion  coupled to a four-form.  In particular, we consider an action of the form
\begin{multline}
 \label{Lorentzianaction1}
S =\int_\mathcal{M} d^4 x\sqrt{|g|} \left[\frac{M_{pl}^2}{2}  R-\frac12  \nabla_\mu \phi \nabla^\mu  \phi -V(\phi) \right] \\+
\int_\mathcal{M} \left[ -\frac{1}{2}F\wedge \star F-\mu \phi  F \right]
 +S_\text{boundary}+S_\text{membranes}
\end{multline}
with $V(\phi)=V_{QFT}-m^4 \cos (\phi/f)$. In terms of our generalised model described earlier, this corresponds to the case where $i, j=1$ only and $\omega^{ij}=1, Z_{ij}=1,\sigma_i=-\mu\phi$.  The bulk constraints  \eqref{con1} to \eqref{con3}  imply that 
\begin{eqnarray}
 3M_{pl}^2 k^2 &=& V_{QFT}-m^4 \cos (\phi/f)+\frac12c^2   \label{con1irr} \\
 \frac{m^4}{f} \sin (\phi/f) &=&\mu c \label{con2irr} \\
\chi &=&-\mu \phi  -c. \label{con3irr}
\end{eqnarray}
It follows that
\be
k^2=\frac{ V_{QFT}-m^4 \cos (\bar \phi(\chi) /f)+\frac{m^8}{2\mu^2 f^2} \sin^2  (\bar \phi(\chi) /f)} {3 M_{pl}^2}\, ,
\ee
where $\bar \phi$ solves the equation
\be
\chi=-\mu \bar \phi-\frac{m^4}{\mu f} \sin (\bar \phi/f)\, . \label{chicond}
\ee
Nucleation of the membrane leads to a jump $\Delta \chi =-Q=-\eta q$, where $q$ is the fundamental membrane charge and $\eta=+1$ for branes and $\eta=-1$ for anti-branes.   This is consistent with the quantisation condition on $\chi=-Nq$ for some $N \in \mathbb{Z}$, with the membrane nucleation giving $\Delta N=\eta$. We can rewrite \eqref{chicond} as
\be
N=\frac{\mu}{q} \bar \phi(N)+\frac{m^4}{q\mu f} \sin (\bar \phi(N)/f)\, , \label{chicond2}
\ee
with the vacuum spectrum characterised by curvature
\be
k^2(N)=\frac{ V_{QFT}-m^4 \cos (\bar \phi(N) /f)+\frac{m^8}{2\mu^2 f^2} \sin^2  (\bar \phi(N) /f)} {3 M_{pl}^2}\, . \label{k}
\ee 
If the cosine potential were generated  by non-perturbative corrections to the axion $\phi$, and nothing else,  we would expect the decay constant and therefore the periodicity of the potential, to be set by the fundamental membrane charge. This corresponds to the critical choice,  $ f_\text{crit}=q/2\pi \mu$. The problem is that in such a set-up, the vacuum curvature \eqref{k} is single valued, making it of no use in addressing the cosmological constant problem.  To see this note that if $f=f_\text{crit}$, then \eqref{chicond} is invariant under $N \to N+n, \bar \phi \to \bar \phi+2\pi n f_\text{crit}$.  It follows that nucleation of a membrane simply triggers a shift in $\bar \phi$ of size $\pm 2\pi  f_\text{crit}$, leaving the curvature unchanged. 

Inspired by the Kaloper-Westphal model, we consider instead the case where $f=w f_\text{crit}$ for some $w\notin \mathbb{Q}$. This could  have arisen from integrating out the heavier of two sectors  in a doubled system with an irrational ratio of membrane charges, reinforcing the connection to Kaloper and Westphal's original set-up.  The curvature spectrum is now dense over a finite range of values.  To see this,  we assume a rational approximation for $w \approx p_1/p_2$, accurate to order $\epsilon$, 
\be
w=\frac{p_1}{p_2}(1+\epsilon)
\ee 
for coprimes $p_1$ and $p_2$.  If $\phi(N)$ solves \eqref{chicond2}, then we can show that  
\be
\phi(N+p_1) \approx \bar \phi (N)+2 \pi  f p_2 -\frac{2\pi f p_1 \epsilon}{\frac{2\pi m^4}{q \mu f}\cos (\bar \phi(N)/f)+w}\, .
\ee
We now compare the vacuum curvature for the vacuum with flux number $N$ and that with number $N+p_1$,
\begin{equation}
\delta k^2=k^2(N+p_1)-k^2(N) \approx-\left[ \frac{2\pi  p_1 m^4 \sin (\bar \phi(N)/f)(1+\frac{m^4}{\mu^2 f^2} \cos (\bar \phi(N)/f)) }{3 M_{pl}^2\left(\frac{2\pi m^4}{q \mu f}\cos (\bar \phi(N)/f)+w\right)}\right] \epsilon.
\end{equation}
Since we can make $\epsilon$ arbitrarily small but non-zero with better and better choices of $p_1$ and $p_2$, it is clear that the distance between vacua can be made arbitrarily small and that the spectrum is dense. It turns out that the curvature lies in a range $k^2 \in [k^2_\text{min}, k^2_\text{max}]$, where
\be
k^2_\text{min}=\frac{1}{3M_{pl}^2}\left(V_{QFT}-m^4\right)
\ee
and
\be \label{kmax}
k^2_\text{max}=
\frac{1}{3 M_{pl}^2}\left[ V_{QFT}+m^4 +\frac12 m^4 \left(\frac{m^2}{\mu f} - \frac{\mu f}{m^2} \right)^2\theta\left(\frac{m^2}{\mu f}-1\right)\right]\ee
where 
\be
\theta\left(\frac{m^2}{\mu f}-1\right)=\begin{cases}
1 & m^2\geq \mu f \\
0 & m^2 < \mu f
\end{cases}.
\ee
We now require that $V_{QFT}$, $\mu f$ and $m^4$ are such that the range of curvature passes from large and positive, $k_{\max} ^2\gg H_0^2 >0$, to large and negative  values, $k_{\min} ^2\ll -H_0^2<0$, with respect to the current Hubble scale. Once this is guaranteed, the density of the spectrum  ensures that vacua exist whose curvature  are sufficiently close to observations. 

Requiring the curvature of vacua to scan either side of the observed value directly constrains the size of the QFT contribution to the vacuum energy. In particular, we find that 
\be
3 M_{pl}^2 H_0^2-m^4 -\frac12 m^4 \left(\frac{m^2}{\mu f} - \frac{\mu f}{m^2} \right)^2\theta \ll V_{QFT} \ll -3 M_{pl}^2 H_0^2+m^4.
\ee
Assuming $m^4 \gg M_{pl}^2 H_0^2$, we can satisfy this just by imposing an upper bound on the scale $|V_{QFT}| \lesssim m^4$. Typically we expect $m^4 \sim \Lambda_\text{UV}^4 e^{-S}$, where $\Lambda_\text{UV}$ is some UV scale\footnote{In QCD, this is set by the QCD scale, or more precisely, by the scale of the quark condensates and the quark masses. In string compactifications it is typically a few orders of magnitude below the Planck scale, being suppressed by the volume of the Calabi-Yau. }  and $S>1$ is the action for the  instanton that generates the potential. Although we cannot extend the scale of the underlying vacuum energy to arbitrarily large values in this model, we can easily accommodate scales just a few orders of magnitude below the Planck scale.

As ever, nucleation of a membrane triggers a jump in one of the flux integers, $N^+ \to N^-=N^+\pm 1$,  translating into a jump in vacuum curvature
\be
\Delta k^2=-\frac{m^4 \Delta [\cos (\bar \phi/f)]}{3 M_{pl}^2} \left[1+\frac{m^4}{\mu^2 f^2} \langle  \cos (\bar \phi/f\rangle)  \right]
\ee
where
\be
\Delta [\cos (\bar \phi/f)]=\cos (\bar \phi(N^+)/f)-\cos (\bar \phi(N^+\pm 1)/f)
\ee
and
\be
\langle \cos (\bar \phi/f)\rangle =\frac12 \left[ \cos (\bar \phi(N^+)/f)+\cos (\bar \phi(N^+\pm 1)/f) \right] \, .
\ee
This depends implicitly on where we are in the landscape through the dependence on $N^+$. Descent from parent de Sitter vacua with curvature $k_+^2 >0$ is characterised by the value of $X=4 M_{pl}^4 \Delta k^2/T^2>0$. If $\tau$ is the fundamental membrane tension, we can use the fact that $\Delta [\cos (\bar \phi/f)]\geq -2$ and $\langle \cos (\bar \phi/f)\rangle\leq 1$ to show that
\be
X\leq \frac{8 M_{pl}^2m^4 }{3  \tau^2} \left[ 1+ \frac{m^4}{\mu^2 f^2}\right]=X_\text{max}\ .
\ee
As in our analysis for the Bousso-Polchinski set-up, dangerous transitions into AdS are kinematically forbidden if $X_\text{max}< X_0(k_+)=4M_{pl}^2 k_+^2/\tau^2$. This introduces a critical curvature $k_c^2=\frac23 \frac{m^4}{M_{pl}^2}\left[ 1+ \frac{m^4}{\mu^2 f^2}\right] $ beyond which we cannnot descened into AdS through a single membrane nucleation. For $k_+^2 \lesssim k_c^2$, descent into AdS may be possible. However, if we impose the modest constraint
\be
\frac{8 M_{pl}^2m^4 }{3  \tau^2} \left[ 1+ \frac{m^4}{\mu^2 f^2}\right]<1 \label{condirr}
\ee
we are guaranteed to have $0<X<1$ for all downward transitions. This ensures that the lifetime of these vacua is determined by a tunnelling exponent  $B\geq B_c$ where 
\be
B_c= \frac{3 M_{pl}^4 \Omega_3}{2m^4}\left[ 1+ \frac{m^4}{\mu^2 f^2}\right]^{-1} \, .
\ee 
By choosing parameters so that this is large we can ensure that the metastable vacua are very long-lived. Furthermore,  the condition \eqref{condirr} also ensures that   
$|X_{M_+ \to AdS_-}|<1$. As we saw previously, this guarantees the presence of a pole in the transition rate as the parent vacuum approaches Minkowski, rendering it arbitrarily long-lived. 

Interestingly, this set-up seems to avoid any issues with the membrane weak gravity conjecture. To see this, note that $f=w f_\text{crit}=\frac{w q}{2 \pi\mu }$,  and so we can rewrite the condition \eqref{condirr} as
\be
\frac{8 M_{pl}^2m^4 }{3  \tau^2} \left[ 1+ \frac{4 \pi^2m^4}{w^2 q^2 }\right]<1 
\ee
which does not indicate any violation of \eqref{wgc}. 
The last thing we need to do is to check that we can indeed transition to a near Minkowski vacuum in a single bungee jump and avoid the empty universe problem.  Using equation \eqref{k}, we see that the flux integer for the near Minkowski vacuum satisfies
\be
\cos \left[\bar \phi\left(N^\text{Mink}\right)/f \right]=-\frac{\mu^2 f^2}{m^4} \left( 1\pm \upsilon \right)
\ee
where 
\be
\upsilon=\frac{1}{\mu f}\sqrt{2\left( V_{QFT}+\frac12 \mu^2 f^2+\frac12 \frac{m^8}{\mu^2 f^2}\right)} \geq \sqrt{6} \frac{M_{pl} k_\text{max}}{ \mu f}
\label{eq:upsilon}
\ee
where in the inequality we have used the formulae for $k_\text{max}$ for both $\mu f \leq m^2 $ and $\mu f >m^2$ given in equation \eqref{kmax}. If Minkowski is the daughter vacuum, it is generated by membrane nucleation in a parent de Sitter vacuum of curvature
\be
k_+^2=\frac{m^4}{3 M_{pl}^2} |\Delta [\cos(\bar \phi/f)]|\left(\upsilon- \frac{m^4}{2 \mu^2 f^2} |\Delta [\cos(\bar \phi/f)]|\right)
\ee
where now
\be
\Delta [\cos(\bar \phi/f)]=\cos(\bar \phi(N^\text{Mink} \mp 1) /f)-\cos(\bar \phi(N^\text{Mink} ) /f)\ .
\ee
Let us now assume that  $\upsilon \gg m^4/ M_{pl}\mu f > H_0$ and $\omega=f/f_\text{crit}>1$. As we will show, these are a set of sufficient conditions that allow us to avoid the empty universe problem.  First up,  note that $\omega>1$ implies  $|\Delta [\cos(\bar \phi/f)]|\sim \mathcal{O}(1)$\footnote{To show that $|\Delta [\cos(\bar \phi/f)]|\sim \mathcal{O}(1)$ for $\omega>1$,  consider the case where $|\Delta [\cos(\bar \phi/f)]| \ll 1$. It follows that $\bar \phi(N^\text{Mink} \mp 1 ) \approx  \sigma \bar \phi(N^\text{Mink})+ 2 n\pi f$ for some $n \in \mathbb{Z}$ and $\sigma=\pm 1$, which is consistent with the condition on the flux integers \eqref{chicond2} if and only if 
\be
\mp 1 \approx n \omega +(\sigma -1)N^\text{Mink} \label{getout}
\ee
 For $\sigma=1$, this implies $\omega \approx 1/|n|$, contradicting the requirement that $\omega>1$. For $\sigma=-1$, the condition \eqref{getout} depends explicitly on $N^\text{Mink}$ and is destabilised by radiative corrections to $V_{QFT}$. We therefore ignore this possibility. }. For $\upsilon \gg \frac{m^4}{\mu^2 f^2}$ and $|\Delta [\cos(\bar \phi/f)]|\sim \mathcal{O}(1)$ the curvature of the parent vacuum is bounded from below as follows
\be
k_+^2\sim \frac{m^4 \upsilon}{3 M_{pl}^2}\ge \frac{m^4 k_{\max}}{M_{pl} \mu f }\, , 
\ee
where in the second inequality we have neglected $\mathcal{O}(1)$ factors and used equation \eqref{eq:upsilon}. As mentioned before, the absence of finely tuned cancellations between the various contributions to the curvature requires  $k_{\max}\gg H_0$. If we further assume that the axion parameters satisfy the inequality $m^4/M_{pl} \mu f> H_0$, as stated above, we  find that
\be
k_+^2\gg \frac{m^4 H_0}{M_{pl} \mu f }\gg H_0^2 .
\ee
Thus, the curvature of the parent vacuum vastly exceeds that of its offspring, the current vacuum. 
It follows that we can bungee jump from a high scale vacuum to the current vacuum with a single bubble nucleation and the empty universe problem is avoided.

We conclude that this model of a single four-form scalar pair  addresses the cosmological constant problem in much the same way as the Kaloper-Westphal model. However, there are some differences. For example, in the  Kaloper-Westphal model, the quantum mechanical stability of Minkowski is independent of $|V_{QFT}|$, whereas here we find it can only be achieved for   $|V_{QFT}|< m^4$. This is not a significant limitation since we can easily imagine $m$ lying just below the Planck scale. This new model also contains certain features that are appealing from the perspective of fundamental theory, including a standard kinetic term for the four-form without any violation of the weak gravity conjecture for membranes.  

Of course, the form of the axion potential may  be further constrained by the {\it axion} weak gravity conjecture \cite{Rudelius:2015xta}. Naively, this would correspond to the condition $f S<M_{pl}$ which is not especially restrictive here. However, we think a more careful analysis  in the context of a particular UV completion of the model would be worthwhile. 

Finally, it is worth noting that besides the vacuum transitions mediated by the nucleation of fundamental membranes charged under the four-form, the present model also allows for transitions between local minima while holding the four-forms fixed. These correspond to the well known Coleman-de Luccia instantons \cite{Coleman:1980aw}, where local vacua are separated by domain walls, and provide an extra channel for the relaxation of the cosmological constant. 

\section{Conclusions} \label{sec:conc}
Within the standard framework of quantum field theory, radiative corrections to the vacuum energy are large, scaling like the fourth power of the cut-off. If we assume naturalness, this  scale should determine the overall size of the renormalized vacuum energy. Corrections to the cosmological constant from other sectors   are now required to bring this result in line with observations, bringing the corresponding vacuum curvature down to the present Hubble scale.  In an effective four-dimensional description of the universe in which gravity is coupled to four-form field strengths, four-form flux provides a correction to the cosmological constant. Nucleation of membranes charged under the corresponding three-form potential, then allows the flux to jump by a single quantised unit, changing the value of the cosmological constant.  This yields a landscape of possible vacua labelled by the quantised flux, scanned through quantum tunnelling. To make contact with observation  the landscape should be  sufficiently dense, with the curvature of vacua separated by no more than the current Hubble scale. Furthermore, to avoid becoming lost in a cold and empty universe, we must be able to reach the current vacuum from a high scale de Sitter vacuum in a single nucleation event. 

In Bousso and Polchinski's seminal paper \cite{Bousso:2000xa}, there are a large number of four-forms,  realising the required density of the landscape without running into the empty universe problem. However, this is not enough to address the cosmological constant problem.  We  also need to ask how the observed vacuum is selected from the landscape of possibilities.  Bousso and Polchinski argue that  the current vacuum should be selected using anthropic arguments: if the vacuum curvature were too large and positive, structure could not have formed due to the rapid exponential expansion; if it were too large and negative, the universe would have started contracting before stars, planets and complex life had ever come into existence. Anthropic arguments have also been used in the context of the electroweak hierarchy problem where, like in the cosmological constant case, naturalness is under pressure. For example,  a coupling  between the Higgs-squared and one \cite{Giudice:2019iwl, Kaloper:2019xfj} or more four-forms \cite{Moretti:2022xlc} yields a landscape of electroweak vacua, from which the current vacuum is selected through anthropics.

In the recent history of string cosmology, anthropic ideas have indubitably become the dominant  explanation for the small value of the cosmological constant seen in nature.  We now seek to challenge that assertion, showing how the current vacuum can often be selected on {\it probabilistic} grounds for a wide class of string-like models of four-forms coupled to scalar fields, including Bousso and Polchinski's original set-up. With this new perspective, our world  exists independently of mankind or some other complex species.  It is born from probability. 

The main idea was inspired by the recent proposal of Kaloper\cite{Kaloper:2022oqv} and  Kaloper and Westphal \cite{Kaloper:2022jpv}, although we have shown how it can be readily extended. In the wide class of models under consideration, we have identified a parameter that controls the lifetime of low scale vacua, similar to the current vacuum we observe. If this parameter can be  suitably bounded, a pole appears in the corresponding bounce actions as the curvature of the parent vacuum goes to zero. This guarantees the quantum stability of the Minkowski vacuum, and the longevity of those vacua close to Minkowski in Planck units.  In the Kaloper-Westphal  model, this bound corresponds to an absolute bound on the membrane charge in terms of its tension. In the Bousso-Polchinski model, the bound is also sensitive to the scale of the renormalised vacuum energy.  However, the sensitivity is not too severe: one can easily accommodate a Planckian energy density for the QFT vacuum, with the membrane tension and charge lying just an order of magnitude or two below the Planck scale. That said, in both the Kaloper-Westphal and the  Bousso-Polchinski set-ups, these probabilistic arguments seem to require a mild violation of the membrane weak gravity conjecture. With this in mind, we have presented a new model where the same probabilistic methods are used but where there is no such violation.

This new approach offers some hope for addressing the cosmological constant problem using probabilistic methods. Notwithstanding some concerns associated with various swampland considerations, this could be relevant to a wide class of four-dimensional effective field theories whose underlying structures are compatible with string compactifications.

\paragraph{Acknowledgements} 
We would like to thank Alex Westphal for useful discussions. AP was supported by STFC consolidated grant number ST/T000732/1 and YL by an STFC studentship.  For the purpose of open access, the authors have applied a CC BY public copyright licence to any Author Accepted Manuscript version arising.

\appendix

\section{Tunnelling rates for general boundary conditions}

For a general choice of boundary condition on the three-forms, the tunnelling exponent computed from the geometry of the bounce is  given by
\be
B=\lambda B_N+(1-\lambda)B_D
\ee
where $B_N$ is the tunnelling exponent computed for Neumann boundary conditions ($\lambda=1$), given by \eqref{BN}, and $B_D$ for Dirichlet boundary conditions, given by
\begin{equation}
B_D=B_N-\Omega_3\sum_i Q_iA^i(0)\\
+\Omega_3 \Delta \left\{\frac{\sum_i \chi_i c^i}{3k^4} \left[3\rho'(r) - \rho'(r)^3\right]^0_{r_\text{min}} \right\}.
\end{equation}

Recall that for Neumann boundary conditions ($\lambda=1$), configurations with $k_+^2 \leq 0$ and $\epsilon_+=-1$ are always infinitely suppressed; for Dirichlet boundary conditions ($\lambda=0$), they will also be infinitely suppressed provided $(2V+\sum_i \sigma_i c^i)_+<0$. If this is not the case, there will be a catastrophic instability. 

For configurations of physical interest, the tunnelling exponent simplifies. Indeed, $B_N$ is now given by \eqref{BNphys} for Neumann boundary conditions on the three-forms, and 
\begin{align}
B_D=& B_N+\Omega_3 \left[ \frac{M_{pl}^2}{3T}  \rho(0)^3 \left(v\Delta k^2 -\frac{T^2}{2 M_{pl}^4} u\right) -\sum_i Q_iA^i(0)\right]\\ \nonumber
&+\frac{2\Omega_3} {3k_+^2 k_-^2}\Bigg\{ (u \Delta k^2-v \langle k^2 \rangle)\left(1- \frac{M_{pl}^2}{T}  \rho(0)\Delta k^2 \right)
\left. +\frac{T}{4M_{pl}^2} \rho(0)\left[ v \Delta k^2  -2 u \langle k^2 \rangle    \right]
\right\}
\end{align}
for Dirichlet boundary conditions, where $u=-6M_{pl}^2+\langle (2V+\sum_i \sigma_i c^i)/k^2 \rangle$ and $v=\Delta \left[(2V+\sum_i \sigma_i c^i)/k^2 \right]$.

\end{document}